\documentclass[aps,prb,floats,showpacs,twocolumn]{revtex4-1}
\usepackage{graphicx,color,wrapfig}
\usepackage{amsmath,amssymb}
\begin{document}
\title{Electron Spin Relaxation in a Transition-Metal Dichalcogenide Quantum Dot} 
\author{Alexander J. Pearce and Guido Burkard}
\affiliation{Department of Physics, University of Konstanz, D-78464 Konstanz, Germany}
%\date{\today}
\begin{abstract}
We study the relaxation of a single electron spin in a circular quantum dot in a transition-metal dichalcogenide monolayer defined by electrostatic gating. Transition-metal dichalcogenides provide an interesting and promising arena for quantum dot nano-structures due to the combination of a band gap, spin-valley physics and strong spin-orbit coupling. First we will discuss which bound state solutions in different B-field regimes can be used as the basis for qubits states. We find that at low B-fields combined spin-valley Kramers qubits to be suitable, while at large magnetic fields pure spin or valley qubits can be envisioned. Then we present a discussion of the relaxation of a single electron spin mediated by electron-phonon interaction via various different relaxation channels. In the low B-field regime we consider the spin-valley Kramers qubits and include impurity mediated valley mixing which will arise in disordered quantum dots. Rashba spin-orbit admixture mechanisms allows for relaxation by in-plane phonons either via the deformation potential or by piezoelectric coupling, additionally direct spin-phonon mechanisms involving out-of-plane phonons give rise to relaxation. We find that the relaxation rates scale as $\propto B^6$ for both in-plane phonons coupling via deformation potential and the piezoelectric effect, while relaxation due to the direct spin-phonon coupling scales independant to B-field to lowest order but scales strongly on device mechanical tension. We will also discuss the relaxation mechanisms for pure spin or valley qubits formed in the large B-field regime.

\end{abstract}  

%\pacs{ }

\maketitle

\section{Introduction}

In recent years two-dimensional semiconductoring monolayers of transition-metal dichalcogenides (TMDs) have become the subject of immense study, due to their intriguing electronic and optical properties.\cite{Wang2012} These atomically thin materials have a direct band gap in the optical frequency range\cite{Mak2010} and a large spin-orbit coupling.\cite{Zhu2011,Kormanyos2015} Their electronic band structure is described by two inequivalent valleys, labeled K and K$'$, and the interplay of this valley degree of freedom and the spin degree of freedom gives rise to rich optical selection rules\cite{Xiao2012,Cao2012,Mak2010a} and coherent manipulation of the valley degree of freedom has been demonstrated.\cite{Jones2013} 

The control of electronic spins states in quantum dot (QD) systems offers a powerful route towards quantum computation.\cite{Loss1998} Tremendous progress has made in the experimental control of electronic spins in GaAs QDs.\cite{Hanson2007} The high density of nuclear spins in GaAs quantum wells has proved to be a limiting factor in their spin decoherence times. This finding has motivated the consideration of other materials as a platforms for QD spin qubits, such as Silicon, carbon nanotubes, and indeed TMDs.

The TMDs are comprised of a honeycomb lattice, with a chemical make up of MX$_2$, in which M is a transition-metal e.g. M=Mo,W and S a chalcogen atom e.g. X=S,Se. Electrostatic gating can be used to create a QD in a semiconducting TMD monolayer. These QDs can also make use of both the valley and spin degrees of freedom to encode information, and also to process this information.\cite{Rohling2012} Recently QDs in TMDs have been experimentally demonstrated both in lateral MoSe$_2$-WSe$_2$ heterojunctions\cite{Huang2014} and in multilayer MoS$_2$, WSe$_2$ and WS$_2$ devices.\cite{Song2015a,Song2015b,Wang2016,Lee2016} Additionally, there have been theoretical studies on the electronic structure of TMD QDs\cite{Kormanyos2014,Liu2014} and their control by optical techniques.\cite{Wu2016} Several experimental groups have also studied the optical properties of confined states formed by crystal vacancies, particularly the excitonic effects in these confining structures.\cite{Tonndorf2015,Srivastava2015a,He2015,Koperski2015,Chakraborty2015}   

The TMDs exhibit a large spin splitting within their band structure near the direct band gap. This creates the situation in QDs, that the level spacing due to confinement is much smaller than the spin splitting $E_L \ll E_{so}$, in stark contrast to GaAs, Silicon and carbon nano-structures. This combined with the valley physics opens up new regimes and opportunities for spin and valley qubits. Indeed, this will form the main focus of this paper in which we will explore the electron spin relaxation time $T_1$ in these TMD QD regimes. The key source of relaxation for electron spin in QDs are interactions with phonons, and phonon excitations in TMDs arise from both the deformation potential and the piezoelectric effect due to the crystals lack of inversion symmetry.\cite{Kaasbjerg2012,Kaasbjerg2013} While in-plane phonons do not direct couple to an electrons spin degree of freedom, they may couple indirectly via a spin-orbit induced mixing of the eigenstates known as the admixture mechanism.\cite{khaetskii2000,khaetskii2001} In contrast, out-of-plane phonons arising due to the TMDs two-dimensional nature, couple directly to the electrons spin via the creation of ripples and curvature with the TMD sheet.\cite{Pearce2016} Together these electron-phonon coupling mechanisms provide a number of differing relaxation channels which are necessary to be understood to ascertain the suitability of TMDs as a platform for future spin-qubits.

The structure of this paper will be as follows: In Sec. II we will introduce a model of the electronic structure of a circular QD in a TMD. Following this in Sec. III we discuss and present scattering rates for spin flip processes caused by phonons in the presence of disorder induced valley mixing, in section III.A those within a Kramers doublet mediated by in-plane acoustic phonons caused by the admixture mechanism while in section III.B we discuss those caused by direct spin-phonon coupling with out-of-plane phonons. Then in Sec. III.C we will consider the regimes in which pure spin or pure valley qubits could be achieved and explore their electron spin relaxation rates. Finally in Sec. IV. we present our conclusion.   

\section{QD Electronic Model}

To study the spin relaxation in a circular QD within a TMD, we assume an electrostatically defined QD in a sample large enough that we may neglect edge effects. The Hamiltonian for the electronic states in the conduction band under the influence of a magnetic field perpendicular to TMD plane is given by
\begin{equation}
H_0 = \frac{\hbar^2}{2m_{\textrm{eff}}} k_{\tau}^{\dagger}k_{\tau} + U(\mathbf{r}) +\lambda \tau_z s_z + \frac{1}{2}g_e\mu_B B_z s_z + \frac{1}{2}g_{v}\mu_{B} B_z \tau_z  \,.
\end{equation}

Here the first term gives the kinetic energy of the electronic states with an electronic momentum $k_{\tau}$ and where $m_{\textrm{eff}}$ is the effective mass in the conduction band, with $m_{\textrm{eff}}/m_e=0.47$ in MoS$_2$ and $m_{\textrm{eff}}/m_e=0.27$ in WS$_2$, here we neglect the small spin and valley dependance of the effective mass.\cite{Kormanyos2014,Kormanyos2015} The electrostatic potential created by electrical gating is captured by $U(\mathbf{r})$, we will assume a potential for a circularly symmetric hard barrier where $U(\mathbf{r}) = U_0\Theta(r-R)$ with $\Theta(r)$ the Heaviside step function, $R$ being the radius of the QD and $U_0$ the height of the confining potential barrier. The third term models the strong spin orbit coupling in TMDs leading to spin splitting in the conduction band, with a magnitude $\lambda$, which is $1.5\textrm{meV}$ in MoS$_2$ while in WS$_2$ it is $-15.5\textrm{meV}$,\cite{Kormanyos2014,Kormanyos2015,Schmidt2016} and the indices $s_z=\pm1$ and $\tau_z=\pm1$ label the spin and valley index respectively. The final two terms describe both the spin and valley Zeeman terms, where $\mu_B$ is the Bohr magneton. The first of these is the well known spin Zeeman term where the effective electron g-factor is given by $g_e=2.21$ in MoS$_2$ and $g_e=2.84$ in WS$_2$. The final term describes a valley Zeeman term which arises as the degeneracy of the valley states is only protected by time reversal symmetry, therefore under the influence of magnetic field the valley degeneracy is broken.\cite{Srivastava2015,Aivazian2015,MacNeill2015} In MoS$_2$ the valley g-factor takes the value $g_v=3.57$, while in WS$_2$ $g_v=4.96$.\cite{Kormanyos2014,Dias2016}   

In the zero magnetic field case we express the momentum as $k_{\tau} = \tau_z k_x+ik_y$. We introduce a magnetic field into the momentum operators with the Kohn-Luttinger prescription, such that $k_i\rightarrow k_i = -i\partial_i+(e/\hbar) A_i$, with $i\in [x,y]$. We choose the symmetric gauge for the vector potential, given by $A=B_z(-y,x)$. Under the influence of a finite magnetic field, the new mechanical momentum operators become non-commuting with the commutation relation $[k_{\tau}^{\dagger},k_{\tau}] = \tau_z 2/l_B^2$, where $l_B$ is the magnetic length which is defined as $l_B=(\hbar/eB_z)^{\frac{1}{2}}$. In radial coordinates we may write the momentum operators as 
\begin{align} k_{\tau}^{\dagger} = & - \frac{i}{l_B} \sqrt{\frac{x}{2}} e^{-i\tau\varphi} \big(\tau_z 2 \partial_x - \frac{i}{x} \partial_{\varphi} + 1\big) \\
k_{\tau} = &  \frac{i}{l_B} \sqrt{\frac{x}{2}} e^{i\tau\varphi} \big( - \tau_z 2 \partial_x - \frac{i}{x} \partial_{\varphi} + 1\big)  \end{align}
where $\varphi$ is the angle measured with respect to the $x$ axis and we have also introduced the dimensionless length scale $x=(r/l_B)^2/2$. 

\begin{figure}
	\centering
		\includegraphics[width=1.0\columnwidth]{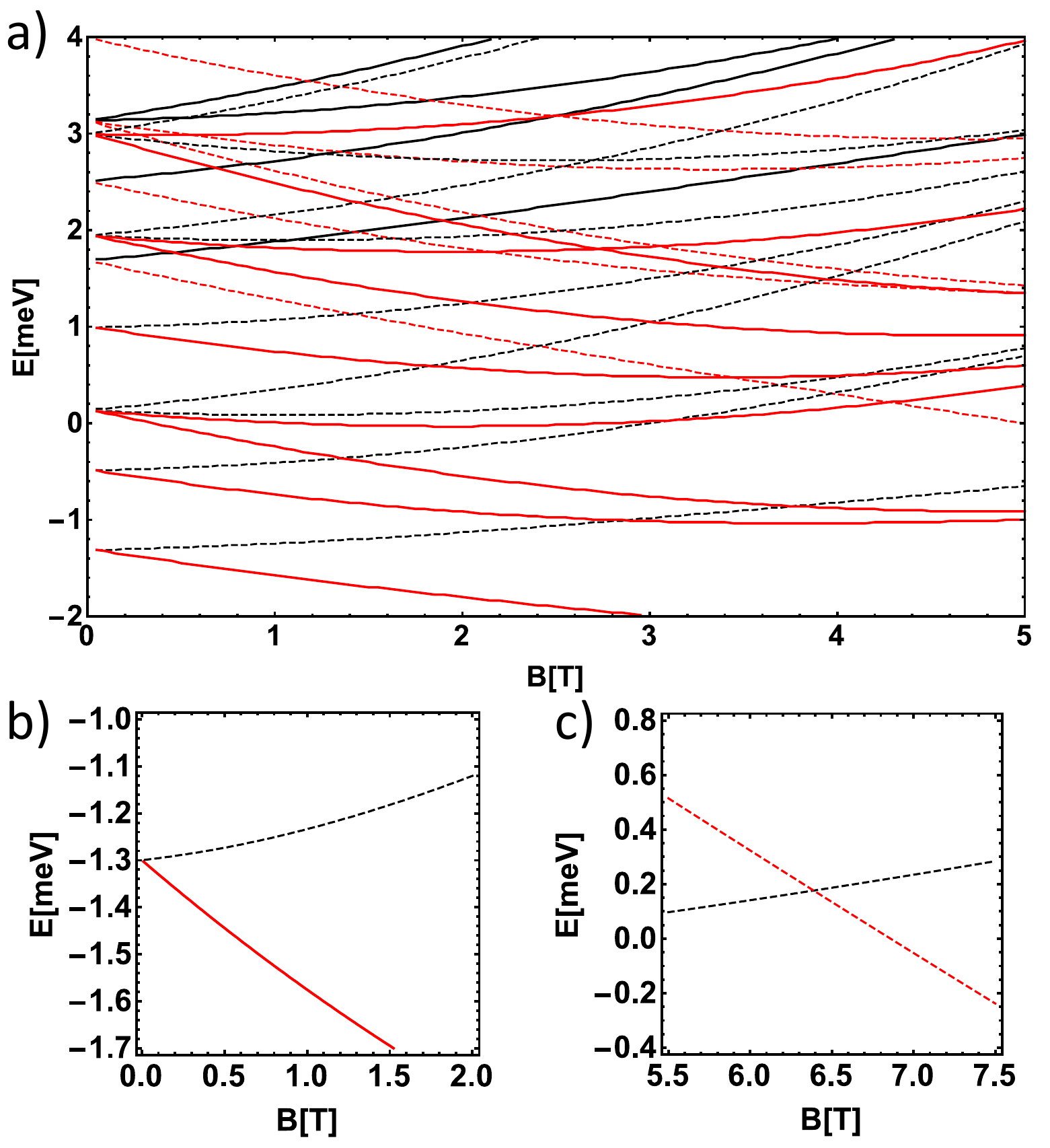}
			\caption{The energy levels in an circular MoS$_2$ quantum dot. Panel (a) shows the electronic spectrum of a QD with a radius $R=50\,\textrm{nm}$ and a barrier heigh of $U_0 = 100\,\textrm{meV}$ as a function of perpendicular magnetic field. Black (Red) lines denote to states with spin up (down) while thick (dashed) lines denote states in the K (K$'$) valley. Here we show the states with orbital and angular momentum quantum numbers $(n,l)$ where $n=0,1,2$ and $l$ satisfying $-n \leq l \leq n$. Panel (b) presents a zoom on the lowest energy Kramers pair states which can serve as a spin-valley qubit in a QD of radius $R=50\,\textrm{nm}$. Panel (c) shows in a zoom around a level crossing between to opposite spin states which both reside within the K$'$ valley at a large magnetic field in a QD of radius $R=20\,\textrm{nm}$.}	
\label{DotLevelsFig}
\end{figure}

Now we seek to solve the eigenvalue equation 
\begin{equation} H_0 \Psi^{\tau}_s(x,\varphi) = E \Psi^{\tau}_s(x,\varphi) \,.\end{equation}
The problem has a cylindrical symmetry and therefore the angular momentum operator $L_z$ commutes with $H_0$ and shares common eigenfunctions. Due to this we factorise the angular component of the solution $\Psi^{\tau}_s(x,\varphi) = e^{il\varphi}\chi_{l,s}^{\tau}(x)$, with $l$ the eigenvalue of $L_z$. The differential equation for $\chi_{l,s}^{\tau}(x)$ is solved by
\begin{align} \chi_{l,s}^{\tau}(x) = &\; x^{\frac{|l|}{2}} e^{-\frac{x}{2}} C^< M(a_{n,l}^{<},|l|+1,x)\Theta(x_0-x) \nonumber \\
& + x^{\frac{|l|}{2}} e^{-\frac{x}{2}} C^> U(a_{n,l}^{>},|l|+1,x)\Theta(x-x_0) \label{eigenstate} \end{align}
where $M(a,b,x)$ and $U(a,b,z)$ are the confluent hypergeometric functions of the first kind and second kind respectively, $C^{\lessgtr}$ are normalisation constants, $x_0=(R/l_B)^2/2$ and $a_{l,\tau}^{\lessgtr}$ is given by the expression 
\begin{align} \hbar \omega_c a_{n,l}^{\lessgtr} =&\, \frac{1}{2} (|l| + l + \tau ) \hbar \omega_c + \lambda_c s_z \tau_z \nonumber \\ 
& + \frac{1}{2}(g_e \mu_B s_z + g_v \mu_B \tau_z)B_z - E^{\lessgtr}_{n,l}  \, . \label{aVal} \end{align}
Here $\omega_c$ is the cyclotron frequency defined as $\omega_c= eB_z/m_{\textrm{eff}}$ and the energies in each region are given by $E^<_{n,l}=\epsilon_{n,l}$ and $E^>_{n,l}=\epsilon_{n,l}-U_0$. Here we consider only $B_z>0$ without loss of generally, as energy levels for negative B-fields are found with the relation $H^{s,\tau}(-B,l) = H^{-s,-\tau}(B,-l).$ Utilising the boundary conditions that demand the matching and continuity of the wavefunction at the QD edge at $x=x_0$ we can acquire the energy levels of the quantum dot. This process yields the characteristic equation
\begin{align} \frac{a_{n,l}^{<}}{|l|+1}M(a_{n,l}^{<}+1,|l|+2,x_0)U(a_{n,l}^{>},|l|+1,x_0) & \nonumber \\
 + a_{n,l}^{>}M(a_{n,l}^{<},|l|+1,x_0)U(a_{n,l}^{>}+1,|l|+2,x_0) & = 0 \;. \end{align}
From this equation we can solve for energy eigenstates for a given spin and valley as a function of B-field, the numerical solution for the energy levels are shown in Fig. \ref{DotLevelsFig}.

%We search for solutions for the following infitatly hard wall problem, where the potential $U(\mathbf{r})$ is given by $U(r) = \Theta()$ step function
%\begin{equation}
%U(\mathbf{r}) = \Bigg  \{ \begin{array}{c}  0 \; \textrm{for} \; r < R \\
%\infty \; \textrm{for} \; r \geq R \end{array}
%\end{equation} 

In the limit of small magnetic field, where $x\ll1$, the confluent hypergeometric function reduce to Bessel functions\cite{AbraStegun,Recher2009} with the relations 
\begin{align} M(a,b,x) = &\; \Gamma (b) (-ax)^{(1-b)/2} J_{b-1} (2\sqrt{-ax}) \\ 
U(a,b,x) = &\; \frac{2}{\Gamma (1+a-b)}  (ax)^{(1-b)/2} K_{b-1}(2\sqrt{xa}) \;, \end{align}
where $\Gamma(n)$ is the gamma function. Using these limits yields an equation for the energy levels of the QD at zero magnetic field, given by 
\begin{figure}
	\centering
		\includegraphics[width=1.0\columnwidth]{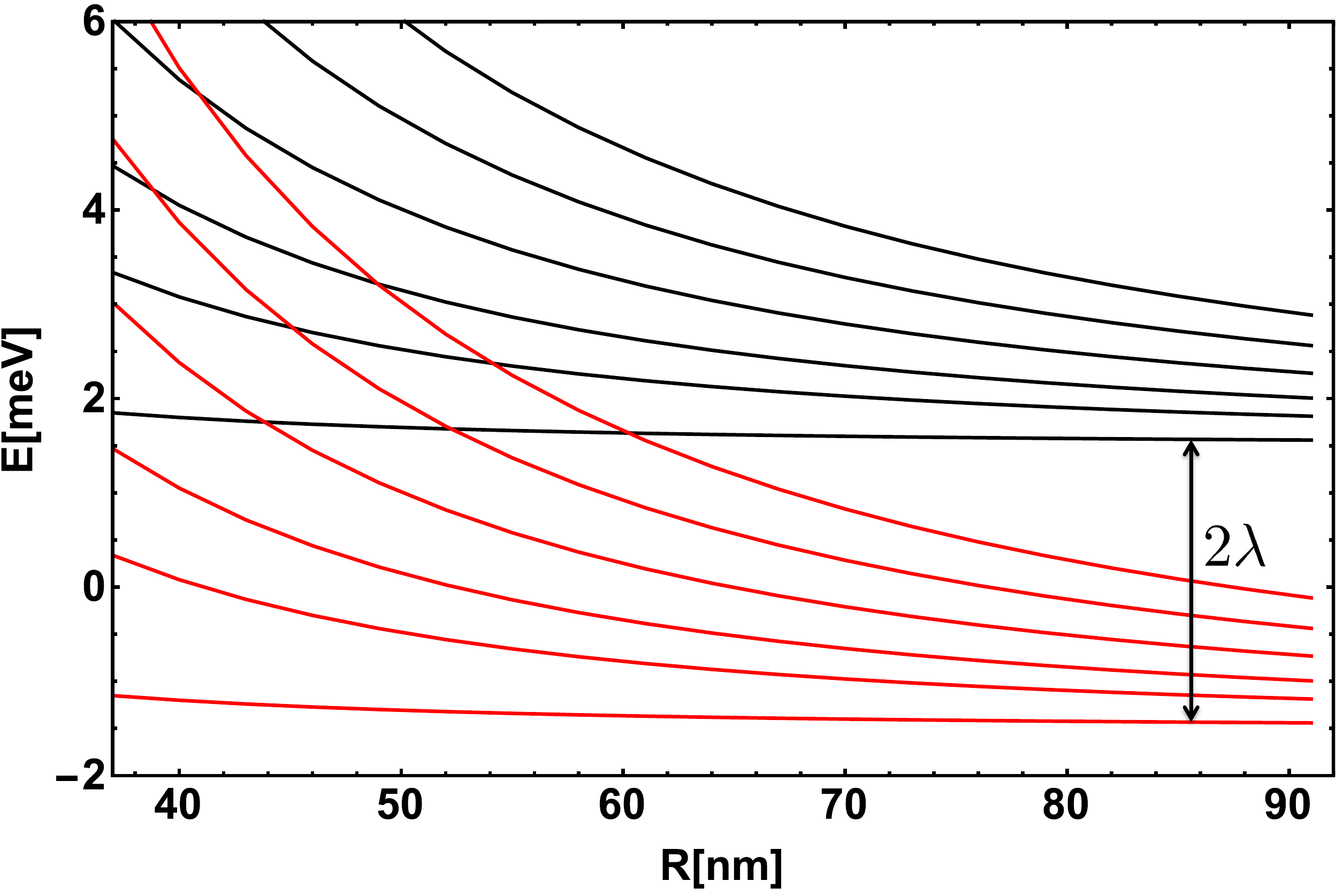}
			\caption{The energy levels in an circular MoS$_2$ quantum dot at zero magnetic field as a function of radius, with a barrier height of $U_0 = 100\,\textrm{meV}$. The red lines denote the states which form a Kramers pair with $(\Psi^{K}_{\uparrow},\Psi^{K'}_{\downarrow})$, while black lines denote the degenerate states with $(\Psi^{K'}_{\uparrow},\Psi^{K}_{\downarrow})$. We plot the lowest energy states, with orbital and angular momentum quantum numbers $(n,l)$ where $n=0,1,2$ and $l$ satisfying $-n \leq l \leq n$. The figure shows an arrow indicating the magnitude of the level splitting $2\lambda$ created by the intrinsic spin-orbit interaction.}	
\label{ZeroFieldDotLevelsFig}
\end{figure}
\begin{align} \sqrt{\xi(1-\xi)}K_{|l|}(2\sqrt{\alpha^>}) J_{|l|+1}(2\sqrt{\alpha^<}) & \nonumber \\
 - K_{|l|+1}(2\sqrt{\alpha^>}) J_{|l|}(2\sqrt{\alpha^<}) & = 0 \;, \end{align}
where $\xi = (\epsilon_{n,l} - \lambda s_z \tau_z)/U_0$, $\alpha^{<} = m_{\textrm{eff}}R^2/2\hbar^2(\epsilon_{n,l} -\lambda \tau_z s_z)$ and $\alpha^{>} = m_{\textrm{eff}}R^2/2\hbar^2(\lambda \tau_z s_z - \epsilon_{n,l} - U_0)$. In Fig. \ref{ZeroFieldDotLevelsFig} we plot the energy levels as a function of dot radius at $B=0\textrm{T}$.

One of the key reasons for interest in QD nanostructure in TMDCs is use as a platform for solid state qubits. We can identify different electronic states which can serve as qubits, these electronic states will depend on the parameter regimes we choose, particularly dot size and magnetic field. 

At zero magnetic field the all electronic states are doubly degenerate Kramers pairs, as seen in both Fig. \ref{DotLevelsFig} and \ref{ZeroFieldDotLevelsFig}, with the pair comprising of the two time reversed partners in each valley, e.g. $(|l,K,\uparrow\rangle,|-l,K',\downarrow\rangle)$. At low magnetic fields where the degeneracy is broken by the spin and valley Zeeman fields these electronic states can used as combined spin-valley qubits, an example of this is presented in Fig. \ref{DotLevelsFig}.b. For example for a dot with $R=50\textrm{nm}$ the energy splitting between the pair of states is found to be $\Delta^{\downarrow\uparrow}_{KK'} \simeq 0.58 \textrm{meVT}^{-1}$, and for instance at 1 Telsa this corresponds to $6.7\textrm{K}$. These states are formed from different valleys and therefore have the advantage that relaxation mechanisms only arise by inter-valley scattering and are robust against small momentum scatterers such smooth disorder or acoustic phonons. In contrast, achieving full control of these qubits can be challenging, although there have been proposals for electron valley resonance in spin-valley qubits.\cite{Palyi2011}

Due to the large spin-orbit coupling in TMDCs, the spin splitting $2\lambda$ is much larger than orbital splitting created due to confinement. Therefore, as a consequence regions of the electronic spectra in which there is a level crossing which would be well suited to pure spin qubits or pure valley qubits are found in the large B-field regime of $B_z>5\,\textrm{T}$. An example of the pure spin qubit in a small radius QD is shown in Fig. \ref{DotLevelsFig}.c.  

\section{Spin Relaxation}

In this section we will study the role of spin relaxation arising due to impurities and interactions with phonons. First, we will focus on the situation of spin relaxation between the energy levels of a Kramers doublet with the degeneracy lifted by Zeeman fields. It is hoped this state could be robust as relaxation necessitates inter-valley scattering, in which impurity scattering would play a crucial role. We consider an impurity driven mixing of the Kramers doublet and in Sec. III.A an admixture mechanism where the electron spin relaxes due to either a piezoelectric phonon or a deformation potential phonon and in Sec. III.B we explore the role of a direct spin-phonon coupling. Additionally in Sec. III.C we will consider the case of a pure spin or valley qubit at a level crossing at large B-field.

Electronic transport and optical studies of TMDs have shown that atomic vacancies and impurities play an important role on the dynamics and relaxation of their charge carriers.\cite{Baugher2013,Yang2015} Indeed, studies of the structure of TMD monolayers have shown high densities of atomic vacancies within the crystal lattice.\cite{Zhou2013,Hong2015,Lin2015} Sharp impurities allow for scattering processes which impart enough momentum transfer to scatter electrons between the differing valleys, K and K$'$, of the Brillouin zone. Here we will consider the role of these inter-valley scattering processes and the limit they may put on $T_1$ times of qubit states formed in a spin-valley space of a Kramers doublet. 

Depending on the particular symmetry properties of an impurity or atomic vacancy the scattering originating from it may be either spin conserving and spin non-conserving. It has been shown that in bulk samples spin non-conserving impurity scattering rates are orders of magnitude smaller than spin conserving scattering rates.\cite{Ochoa2014,Habe2016} Therefore in this work we will focus on spin conserving inter-valley scattering. However combined with the presence of spin-orbit fields this disorder will lead to spin relaxation mechanisms. Here will always assume that we are at zero temperature, this will continue be a good approximation given that $k_BT \ll g\mu_BB_z$.%Therefore we see that, if we wish to operate our qubit within some small subsection of the QD states understanding these inter-valley scattering mechanisms is crucial. In the presence of spin-orbit interactions, .When considering spin relaxation between states in the same valley, as we did in the previous section, impurity mediated spin relaxation is not dominant WHY?.Another regime of interest for electronic states which could be used as a qubit is the low B-field regime. In this regime the states of interest are of opposite valleys and spins and related by time reversal symmetry and form a Kramers pair at zero B-field. It is hoped this state could be robust as relaxation necessitates inter-valley scattering, in which impurity scattering would play a crucial role.   

To model the short range disorder within the crystal lattice we introduce into the Hamiltonian a term which mixes the valleys,\cite{Ando1998,Palyi2010} given by $H_{\textrm{dis}} = \tau_x V_{KK'}/2$. We now consider a QD eigenstate of $H_0$, given by $\langle r,\varphi | n, \tau, s\rangle = \Psi^{\tau}_s(r,\varphi)$ with $n$ labelling an orbital state and $\tau$ and $s$ the valley and spin respectively, and perturb it by the valley mixing term $H_{\textrm{dis}}$. We find this short range disorder weakly mixes the valley states as,
\begin{align} |\tilde{K}\rangle = & \frac{1}{\sqrt{1+\gamma^2_s}}\big(|K\rangle^{(0)} - \gamma_s |K'\rangle^{(0)} \big) \nonumber \\ 
|\tilde{K}'\rangle = & \frac{1}{\sqrt{1+\gamma^2_s}}\big( \gamma_s |K\rangle^{(0)} + |K'\rangle^{(0)} \big) \;, \end{align}
where $\gamma_s = V_{KK'}/[\Delta\epsilon_s + \sqrt{\Delta\epsilon_s^2+V_{KK'}^2}]$ with $\Delta\epsilon_s = \epsilon_{K,s}-\epsilon_{K',s}$ in which $s=\uparrow,\downarrow$ and the superscript $(0)$ refers to an unperturbed state. These energies can be calculated from the solutions to Eq. (\ref{aVal}) and we note that the mixing of the valley states therefore depends indirectly on the spin via these energies $\Delta \epsilon_s$. 

The matrix element for the relaxation of a spin between a Kramers doublet formed in a orbital level $n$ by the electron-phonon Hamiltonian $H^{\alpha}_{\textrm{el-ph}}$ causing the with the emission of a phonon of type $\alpha$ can be now written in the unperturbed valley eigenstates, this process yields  
\begin{multline} \langle n , \tilde{K}' , \downarrow | H^{\alpha}_{\textrm{el-ph}} | n , \tilde{K} , \uparrow \rangle = \nu_{\downarrow} \langle n , K , \downarrow  | H^{\alpha}_{\textrm{el-ph}} | n , K, \uparrow \rangle \\
 - \nu_{\uparrow} \langle n , K' , \downarrow | H^{\alpha}_{\textrm{el-ph}} | n , K' , \uparrow \rangle \;, \label{spinflipME} \end{multline}
where
\begin{equation} \nu_s = \frac{\gamma_s}{\sqrt{(1+\gamma_{\uparrow}^2)(1+\gamma_{\downarrow}^2)}} \;. \end{equation}
It is instructive to consider the limiting case of $\nu_s$, in which $V_{KK'} \ll \Delta\epsilon_s$ which holds for any reasonable energy scale of inter-valley mixing. In this case the prefactor reduces to $\nu_{s} = V_{KK'}\Delta\epsilon_{s}/(2\Delta\epsilon_{\downarrow}\Delta\epsilon_{\uparrow})$. We note that $\nu_s$ depends weakly on B-field and for zero B-field it is given by the simple relation $\nu_{\downarrow}(B_z=0) = - \nu_{\uparrow}(B_z=0) = V_{KK'}/4\lambda$.

We find then that impurities and vacancies mix the valley eigenstates allowing for spin relaxation within a Kramers doublet, but the matrix element is suppressed by the factors $\nu_s$. It now becomes imperative to evaluate spin flip matrix element in Eq. (\ref{spinflipME}), which we will consider in detail in the following two sections.

%To compute the full relaxation time we use the Fermi golden rule, given as
%\begin{equation} \frac{1}{T_1} = \Gamma_{\alpha} = \frac{2\pi A}{\hbar} \int \frac{d^2q}{(2\pi)^2} |\langle n , \tilde{K}' , \downarrow | H^{\alpha}_{\textrm{el-ph}} | n , \tilde{K} , \uparrow \rangle |^2 \delta (\hbar v^{\alpha}q - g_e \mu_B B) \;. \label{rateEq} \end{equation}
%where $\alpha = \phantom{}^{(0)}\langle K' |V_{KK'}| K \rangle^{(0)} / (\tilde{g}\mu_B B)$, where $\tilde{g}=g_e+g_v$. 
%we describe the role of disorder with $H_{\textrm{dis}} = \sum_i V_{0,i} \delta(\mathbf{r} - \mathbf{r}_i)$, where $V_{0,i}$ is a random potential due to an impurity found at a position $\mathbf{r}_i$. 
%As a consequence we will focus our attention only on spin-conserving scattering processes. 
%as that the presence of disorder will mix the valley states,
%\begin{equation} V_{KK'} = (\delta_x \tau_x + \delta_y \tau_y) \;, \end{equation}%\otimes s_0
%where $\delta_{x,y}$ gives the disorder strength. TO THIS POINT.
%Include details...

\subsection{Admixture Mechanism}

The admixture mechanism refers to the effect caused by a Rashba type spin-orbit interaction which weakly mixes the eigenstates in Eq. (\ref{eigenstate}), allowing for phonons to couple to the election spin indirectly causing a relaxation mechanism.\cite{khaetskii2000,khaetskii2001} We now consider the matrix element found in Eq. (\ref{spinflipME}) and its QD eigenstates of $H_0$, given by $\langle r,\varphi | n, \tau, s\rangle = \Psi^{\tau}_s(r,\varphi)$ with $n$ labelling an orbital state and $s$ the spin. This eigenstate is perturbed by the induced spin-orbit interaction. In lowest order in $H_{\textrm{so}}$ the perturbed eigenstate is given by 
\begin{equation} | n \uparrow \rangle = | n \uparrow \rangle^{(0)} + \sum_{k\neq n} | k \downarrow \rangle^{(0)} \frac{\phantom{}^{(0)}\langle k \downarrow | H_{\textrm{so}} | n \uparrow \rangle^{(0)}}{\epsilon_{n}-\epsilon_{k}}\,, \end{equation}
here once more the superscript $(0)$ refers to an unperturbed state. %The energy $E_n$ is the We now consider eigenstates with in the same valley but opposite spins which cross at a finite B-field $B_0$, and study the relaxation of the spin within the same valley and orbital level
Therefore the matrix element for a spin-flip transition within an orbital level $n$ of given valley $\tau$ with the emission of a phonon of type $\alpha$ is given by 
\begin{align} \langle n , \tau , \downarrow | H^{\alpha}_{\textrm{el-ph}} | n , \tau , \uparrow \rangle = &\, \sum_{k\neq n} \Big[\frac{(H^{\alpha}_{\textrm{el-ph}})_{nk}^{\tau\tau}(H_{\textrm{so}})_{kn}^{\downarrow\uparrow,\tau\tau}}{E^{\tau}_{n}-E^{\tau}_{k}+\tau_z2\lambda-g_e\mu_BB_z} \label{phmatel} \nonumber \\ 
+ &\, \frac{(H_{\textrm{so}})_{nk}^{\downarrow\uparrow,\tau\tau}(H^{\alpha}_{\textrm{el-ph}})_{kn}^{\tau\tau}}{E^{\tau}_{n}-E^{\tau}_{k}-\tau_z2\lambda+g_e\mu_BB_z} \Big] \,. \end{align}
Where here the notation $(H)_{ij}^{ss',\tau\tau'}$ is short hand for $\phantom{}^{(0)}\langle i \tau s | H | j \tau' s' \rangle^{(0)}$, and $E^{\tau}_n$ refers to the non-spin dependant terms within the orbital energy eigenstates.% and typical values for $B_0\simeq4\textrm{T}$.

At low B-fields the we assume the dominant relaxation will be between the electronic states which form Kramers doublets, i.e. at zero B-field they are time reversed partners. If the two eigenstates are related by time reversal, then we note that $(H_{\textrm{so}})_{kn}^{\downarrow\uparrow,KK} = -(H_{\textrm{so}})_{nk}^{\downarrow\uparrow,K'K'}$ and as a consequence the matrix element shown in Eq. (\ref{phmatel}) will be matched by the process in the opposite valley and the relaxation mechanism will vanish in the absence of a magnetic field, this effect is known as Van Vleck cancelation,\cite{Vleck1940,Abrahams1957} and leads to a higher order dependance on the B-field of the matrix element. As we assume small B-fields we therefore expand the matrix element in the small field limit $g_e\mu_BB_z/(\epsilon_n-\epsilon_k)\ll1$.

We examine the effect of a Bychkov-Rashba type spin-orbit interaction, which is given by the Hamiltonian
\begin{equation} H_{\textrm{so}} =  \lambda_{R} k_{\tau} s^- + \lambda_{R}^* k_{\tau}^{\dagger} s^+ \,, \end{equation}   
where $s^{\pm}$ are spin raising and lower operators and $\lambda_{R}$ is the Bychkov-Rashba coupling constant. This spin-orbit interaction term arises when the mirror inversion symmetry in the $z$ direction is broken, this can by ripples and curvature of the two-dimensional sheet\cite{Pearce2016} or by the application of a perpendicular electric electric field. In the case of a perpendicular electric field $E_z$ the Bychkov-Rashba coupling constant has been predicted to be $|\lambda_R| = 0.033E_z[\textrm{V}/\textrm{\AA}]\textrm{eV\AA}$ in MoS$_2$ and $|\lambda_R| = 0.13E_z[\textrm{V}/\textrm{\AA}]\textrm{eV\AA}$ in WS$_2$.\cite{Kormanyos2014}

We calculate the matrix element for spin-orbit interaction between states of differing orbital states. This yields
\begin{equation} (H_{\textrm{so}})_{nk}^{\downarrow\uparrow,\tau\tau} = i\frac{2\sqrt{2}\pi\lambda_R}{l_B} \delta_{n,k+\tau} (k_{\tau})_{nk}^{\tau\tau} \;, \end{equation}
where the orbital part of the matrix element $(k_{\tau})_{nk}^{\tau\tau}$ depends on initial angular momentum quantum number and the full details of the momentum operator is discussed in greater depth in Appendix A. Evaluation of momentum overlap gives
\begin{align} (k_{\tau})_{nk}^{\tau\tau'} = &\; - \tau'k\Theta(-\tau'k)M^{\tau\tau'}_{n,k-1} \nonumber \\
& \; + \Theta(\tau'k)\Big[\frac{\Theta(\tau')-\tau' a^{<}_{n,k}}{|k|+1}\Big]M^{\tau,\tau'}_{n,k+1} \;, \end{align}
where $\Theta(x)$ is the Heaviside step function and $M^{\tau,\tau'}_{n,n'} = \int dx x \chi_{n,s}^{\tau *}(x) \chi_{n',s}^{\tau'}(x)$.

The Hamiltonian $H_{\textrm{el-ph}}^{\alpha}$ describes the electron phonon interaction with a phonon of type $\alpha$ and with wave vector $\mathbf{q}$. Here we consider both electron phonon coupling arising due to the piezoelectric interaction and the deformation potential, and as we are interested in low energy regime we consider only the acoustic phonon modes. The deformation potential (DP) arises due to local variations of area of the TMD sheet and therefore couples only to longitudinal (L) phonon modes. Whereas, the piezoelectric effect (PE) creates a electron phonon interaction which couples to both longitudinal (L) and transverse (T) phonon modes. 

In bulk 3D materials which support an electron-phonon interaction with piezoelectric phonons the coupling is known to be independent of the phonon wavenumber $\mathbf{q}$, but in contrast, in 2D materials in the long wavelength limit the electron-phonon interaction is linear in $\mathbf{q}$.\cite{Kaasbjerg2013} This means that the phonon wavenumber dependance of the electron-phonon coupling is the same for both deformation and piezoelectric phonons and they only differ by details of their coefficients. Therefore these three separate Hamiltonians for these distinct electron phonon interaction channels have the similar forms of
\begin{equation} H_{\textrm{el-ph}}^{\alpha} = |\mathbf{q}| D^{\alpha}(\phi_{\mathbf{q}})  \zeta_{\mathbf{q}}^{(\alpha)} (e^{i\mathbf{q}\cdot\mathbf{r}}a^{(L)}_{\mathbf{q}} + e^{-i\mathbf{q}\cdot\mathbf{r}}a^{(L)\dagger}_{-\mathbf{q}} ) \;, \end{equation}
%\begin{align} H_{\textrm{el-ph}}^{\alpha=\textrm{DP,L}} = &\, i g_1 |\mathbf{q}| \zeta_{\mathbf{q}}^{(L)} (e^{i\mathbf{q}\cdot\mathbf{r}}a^{(L)}_{\mathbf{q}} + e^{-i\mathbf{q}\cdot\mathbf{r}}a^{(L)\dagger}_{-\mathbf{q}} ) \\
%H_{\textrm{el-ph}}^{\alpha=\textrm{PE,L}} = &\, - g_2 \zeta_{\mathbf{q}}^{(L)} \cos3\phi (e^{i\mathbf{q}\cdot\mathbf{r}}a^{(L)}_{\mathbf{q}} + e^{-i\mathbf{q}\cdot\mathbf{r}}a^{(L)\dagger}_{-\mathbf{q}} ) \\
%H_{\textrm{el-ph}}^{\alpha=\textrm{PE,T}} = &\, g_2 \zeta_{\mathbf{q}}^{(T)} \sin3\phi  (e^{i\mathbf{q}\cdot\mathbf{r}}a^{(T)}_{\mathbf{q}} + e^{-i\mathbf{q}\cdot\mathbf{r}}a^{(T)\dagger}_{-\mathbf{q}} ) \,.\end{align}
where coefficient for the deformation potential is given by $D^{\textrm{DP,L}}(\phi_{\mathbf{q}}) = i g_1$, with $g_1=2.4\textrm{eV}$\cite{Kaasbjerg2012}, whereas the coefficients for the piezoelectric interaction are $D^{\textrm{PE,L}}(\phi_{\mathbf{q}}) = - [e_{xx} e /\epsilon_0] \cos3\phi_{\mathbf{q}}$ and $D^{\textrm{PE,T}}(\phi_{\mathbf{q}}) = [e_{xx} e /\epsilon_0] \sin3\phi_{\mathbf{q}}$ in which the piezoelectric interaction is anisotropic in $\mathbf{q}$, and the angle of the phonon is measured by $\tan\phi_{\mathbf{q}}=q_y/q_x$, and the piezoelectric constant is $e_{xx}= 3.0 \times10^{-11}\textrm{C/m}$.\cite{Kaasbjerg2013} Here the oscillator length is given by $\zeta_{\mathbf{q}}^{(\alpha)} = [\hbar/2\rho A \omega_{\mathbf{q}}^{(\alpha)}]^{\frac{1}{2}}$, with $\rho$ the mass density of the TMDC sheet and $A$ the area. The in-plane longitudinal and transverse dispersions are linear and are given by $\omega_{\mathbf{q}}^{(L)} = v^{(L)}\mathbf{q}$ and $\omega_{\mathbf{q}}^{(T)} = v^{(T)} \mathbf{q}$. In MoS$_2$ and the group velocities have been calculated with first principles techniques as $v^{(L)}= [(\lambda+2\mu)/\rho]^{\frac{1}{2}}\simeq 6.7\times10^3\textrm{m/s}$ and $v^{(T)}=[\mu/\rho]^{\frac{1}{2}}\simeq 4.2\times10^3\textrm{m/s}$.\cite{Kaasbjerg2012,Kaasbjerg2013} We consider that for the case of our interest, where $\hbar v^{\alpha}\mathbf{q}=g_e\mu_BB_z$ we find that the phonon wave length will be approximately $\lambda \simeq 200\,\textrm{nm}$ for $B_z=1\textrm{T}$, which is much larger that typical size of QDs and justifies the use of the dipole approximation within our work.

The electron-phonon matrix element for the electron phonon interaction arising from a phonon of type $\alpha$ is found to be 
\begin{multline} (H^{\alpha}_{\textrm{el-ph}})_{nk} = i2\sqrt{2}\pi D^{\alpha}(\phi_{\mathbf{q}}) l_B |\mathbf{q}|^{\frac{3}{2}} \sqrt{\frac{\hbar}{2A\rho v^{(\alpha)}}} \nonumber \\
\times N_{n,k} (\delta_{n,k-1}e^{i\phi_{\mathbf{q}}} +\delta_{n,k+1}e^{-i\phi_{\mathbf{q}}})\;. \end{multline}
where $N^{\tau,\tau'}_{n,n'} = \int dx x^{\frac{3}{2}} \chi_{n,s}^{\tau *}(x) \chi_{n',s}^{\tau'}(x)$. %Whereas the matrix elements for the piezoelectric electron-phonon interactions are given by 
%\begin{equation} (H^{\textrm{L},\textrm{PE}}_{\textrm{el-ph}})_{nk} = D_{\textrm{L}}  |\mathbf{q}|^{\frac{1}{2}}\cos 3\phi_{\mathbf{q}} N_{n,k} (\delta_{n,k-1}e^{i\phi_{\mathbf{q}}} +\delta_{n,k+1}e^{-i\phi_{\mathbf{q}}})\;. \end{equation}
%\begin{equation} (H^{\textrm{T},\textrm{PE}}_{\textrm{el-ph}})_{nk} = D_{\textrm{T}} |\mathbf{q}|^{\frac{1}{2}}\sin 3\phi_{\mathbf{q}} N_{n,k} (\delta_{n,k-1}e^{i\phi_{\mathbf{q}}} +\delta_{n,k+1}e^{-i\phi_{\mathbf{q}}})\;. \end{equation}
%where $D_{\alpha}= -i 2\pi g_2 [\hbar/2A\rho v^{(\alpha)}]^{\frac{1}{2}}$ and $\alpha=\textrm{L},\textrm{T}$ refers to either the longitudinal and transverse phonon modes.

To compute the full relaxation time we use Fermi's golden rule, given as
\begin{equation} \Gamma_{\textrm{Total}} = \frac{1}{T_1} = \sum_{\alpha} \Gamma_{\alpha}\; , \end{equation}
\begin{equation} \Gamma_{\alpha} = \; \frac{2\pi A}{\hbar} \int \frac{d^2q}{(2\pi)^2} |\langle n \uparrow | H^{\alpha}_{\textrm{el-ph}} | n \downarrow \rangle |^2 \delta (\hbar v^{(\alpha)}q - \tilde{g} \mu_B B_z) \;, \label{rateEq} \end{equation}
using the definition $\tilde{g} = g_e + g_v$. Using all the ingredients we have discussed until now, we are in position to calculate the relaxation rates. First we consider the electron spin relaxation arising from interactions with longitudinal  deformation potential phonons and find the relaxation rate 
\begin{align} & \Gamma_{\textrm{L},\textrm{DP}} = \; (\nu_{\uparrow} + \nu_{\downarrow})^2 \frac{16\pi^4\lambda^2_{R}g_1^2}{\rho v^{(L)6} \hbar^5} g_e^2\tilde{g}^4\mu_B^6B_z^6 \nonumber \\
& \times \Bigg| \sum_{k\neq n} \delta_{n,k+1}\frac{[(k_K)^{K,K}_{n,k}N^{K,K}_{k,n}+N^{K',K'}_{n,k}(k_{K'})^{K',K'}_{k,n}]}{(E^{K'}_n - E^{K'}_k - 2\lambda)^2} \nonumber \\
& - \delta_{n,k-1} \frac{[N^{K,K}_{n,k}(k_K)^{K,K}_{k,n}+(k_{K'})^{K',K'}_{n,k}N^{K',K'}_{k,n}]}{(E^{K}_n - E^{K}_k + 2\lambda)^2} \Bigg|^2 \;. \label{RelaxDefPot} \end{align}
In two-dimensional materials long wavelength the deformation potential and piezoelectric electron-phonon coupling has the same $\mathbf{q}$ dependance, and therefore the relaxation rates due to piezoelectric phonons will have the same dependance on B-field. As a consequence the relaxation rates by either longitudinal and transverse phonon modes are found to only differ by coupling parameters and group velocity from deformation potential phonon mediated relaxation. These relaxation rates are given by
\begin{align} \Gamma_{\textrm{L},\textrm{PE}} = &\; \frac{e^2e_{xx}^2}{2g_1^2\epsilon^2_0} \Gamma_{\textrm{L},\textrm{DP}} \label{RelaxPiezo2} \\
\Gamma_{\textrm{T},\textrm{PE}} = &\; \frac{e^2e_{xx}^2v^{(L)6}}{2g_1^2\epsilon^2_0 v^{(T)6}} \Gamma_{\textrm{L},\textrm{DP}} \label{RelaxPiezo} \end{align}
%\begin{align} \Gamma_{\alpha,\textrm{PE}} = &\; \frac{2\pi^2\lambda^2_{BR}m^2_{\textrm{eff}}g_2^2}{\rho v^{(\alpha)4}\hbar^7} [g_e\mu_B(B-B_0)]^2 \nonumber \\
%& \times \Big| N^{a,a}_{n,n+1}N^{a,a}_{n+1,n} - N^{a,a}_{n-1,n}N^{a,a}_{n,n-1} \Big|^2 \;, \end{align}
\begin{figure}
	\centering
		\includegraphics[width=1.0\columnwidth]{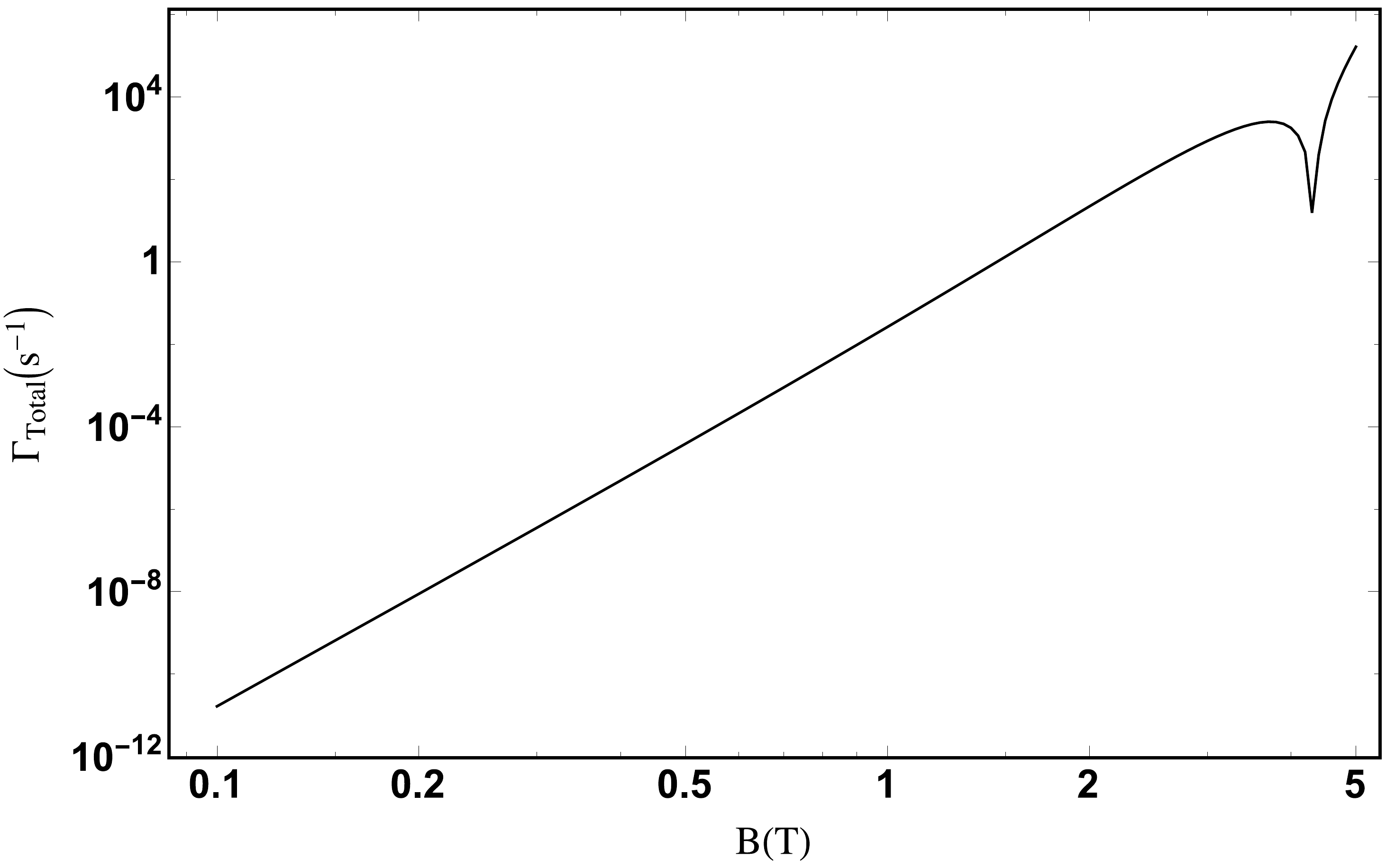}
			\caption{The relaxation rate $\Gamma$ between Kramers pair states $(\Psi^{K}_{\uparrow},\Psi^{K'}_{\downarrow})$ as a function of perpendicular magnetic field. The plotted line is the total relaxation rate given by $\Gamma_{\textrm{Total}} = \Gamma_{\textrm{L},\textrm{DP}} + \Gamma_{\textrm{L},\textrm{PE}} + \Gamma_{\textrm{T},\textrm{PE}}$. The calculation is preformed for circular MoS$_2$ quantum dot with a radius of $R=50\textrm{nm}$ with hard wall boundary conditions and with a inter-valley coupling of $V_{KK'} = 0.1\textrm{meV}$. We consider the relaxation within the lowest energy Kramers pair and sum over all other possible intermediate states.}	
\label{RelaxationRateFig}
\end{figure}

We find that the relaxation due to phonons arising from both deformation potential and piezoelectric effect varies with the sixth power of the perpendicular magnetic field. A plot the relaxation rates due to the admixture mechanism in Eq. (\ref{RelaxDefPot})-(\ref{RelaxPiezo}) are presented in Fig. \ref{RelaxationRateFig}. We see that the need for inter-valley mixing by disorder leads to a suppression in the relaxation rate and increase in the $T_1$ time. The dominant contribution to $T_1$ arises from transverse piezoelectric phonons relaxation rate $\Gamma_{\textrm{T},\textrm{PE}}$ as these phonons possess the most efficient combination of electron-phonon coupling and phonon density of states. Additionally in Fig. \ref{RelaxationRateFig} we note a non-monotonic behaviour in the relaxation rate at a magnetic field of approximately $3.8\textrm{T}$, this arises due to a destructive interference of matrix elements of scattering processes in different valleys. Therefore this behaviour is only possible for relaxation between states of different spin-valley states and will be characteristic of a Kramers qubit which is weakly mixed by disorder. This destructive valley interference is maximised at a magnetic field corresponding to a level crossing between intermediate states in opposite valleys. While we would expect to see this non-monotonic behaviour in all TMD Kramers qubits, the exact magnetic field in which they would occur in other TMD, such as WS$_2$, would depend on material parameters. 

We can contrast this behaviour with studies within other materials. Most notably GaAs QDs have a predicted $B^5$ dependance in their spin relaxation rates and while it also exhibits a Van Vleck cancelation the bulk three-dimensional electron phonon coupling for piezoelectric phonons leads to the differing dependance on B-field.\cite{khaetskii2001} Non-monotonic behaviour has also been predicted in the $T_1$ of carbon nanotube quantum dots, but this is predicted due to interference between discrete and continuum electronic states.\cite{Bulaev2008} In addition, in closer relation to TMDs, the spin relaxation in graphene QDs for an admixture mechanism mediated by deformation potential phonons has a $B^4$ predicted dependance, due to a lack Van Vleck cancelation when considering only one valley degree of freedom.\cite{Struck2010} Indeed, within a later section of this work we will consider the parameter regime when analogous physics can also be created in TMD QDs. 

\subsection{Direct Spin-Phonon Coupling}

We now turn our attention to the direct spin-phonon processes. The coupling between phonons and electronic spins arises due to flexural phonons, these out-of-plane phonons break the mirror symmetry of the lattice allowing for finite transition rates for spin flip processes. There are two mechanisms which we discuss in this section. The first is a deflection coupling and is a geometric effect arising due to local rotation of the lattice created out-of-plane deformations,\cite{Struck2010,Rudner2010} while the second is due to local curvature of the crystal lattice which mixes the electronic orbitals such that non-spin conserving processes are allowed.\cite{Pearce2016} These mechanisms are distinct as tilting the crystal lattice does not change the electronic structure of the TMD.

The deflection coupling is caused by a long wavelength acoustic flexural phonons which induces a local tilt in two-dimensional crystal lattice and the electronic spin is defined in the local reference frame with respect to the local normal vector $\hat{\mathbf{z}}'$ as $s_z = \mathbf{s}\cdot\hat{\mathbf{z}}'$. Therefore under local tilts of the crystal lattice the spin-orbit splitting contained in $H_0$ within the conduction band, which in the flat case is given by $H_{\textrm{so},0} = \lambda\tau_z s_z$, becomes
\begin{equation} H_{\textrm{so},0} = \lambda\tau_z s_z - \lambda\tau_z (s_x\partial_x h(\mathbf{r})+s_y\partial_y h(\mathbf{r}))\;, \end{equation}
and now depends on the local tilt of lattice. In an effort the study the coupling with phonons we quantise the high deformation field as $h(\mathbf{r}) = \sum_{\mathbf{q}} \zeta^{(F)}_{\mathbf{q}} \exp[i\mathbf{q}\cdot\mathbf{r}] (a^{(F)}_{\mathbf{q}} + a^{(F)\dagger}_{-\mathbf{q}})$, where $a^{(F)}_{\mathbf{q}}$($a^{(F)\dagger}_{\mathbf{q}}$) is an operator which creates (annihilates) a flexural phonon with a wave vector $\mathbf{q}$. 

The dispersion relation for flexural phonons is $\omega^{(F)}_{\mathbf{q}} = [(\kappa|\mathbf{q}|^4+\gamma|\mathbf{q}|^2)/\rho]^{\frac{1}{2}}$, where here $\kappa=9.61\textrm{eV}$\cite{Jiang2013} and is the energy cost for bending the TMDC lattice and $\Gamma$ is a sample specific surface tension induced when the TMD monolayer is contacted in a realistic device design which breaks the rotational symmetry of the membrane. The out-of-plane flexural dispersion is described by two regimes governed by the wave vector scale $q_{*}=[\gamma/\kappa]^{1/2}$, above this scale the dispersion is quadratic with $\omega^{(F)}_{\mathbf{q}}=\beta_{\textrm{LT}}|\mathbf{q}|^2$ with $\beta_{\textrm{LT}}=[\kappa/\rho]^{\frac{1}{2}}$ and below it is linear with the dispersion $\omega^{(F)}_{\mathbf{q}}=\beta_{\textrm{HT}}|\mathbf{q}|$ with $\beta_{\textrm{HT}}=[\gamma/\rho]^{\frac{1}{2}}$, so we see that sample dependant tension due to clamping stiffens the flexural modes and decreases their density of states at low wave vectors. 

\begin{figure}
	\centering
		\includegraphics[width=1.0\columnwidth]{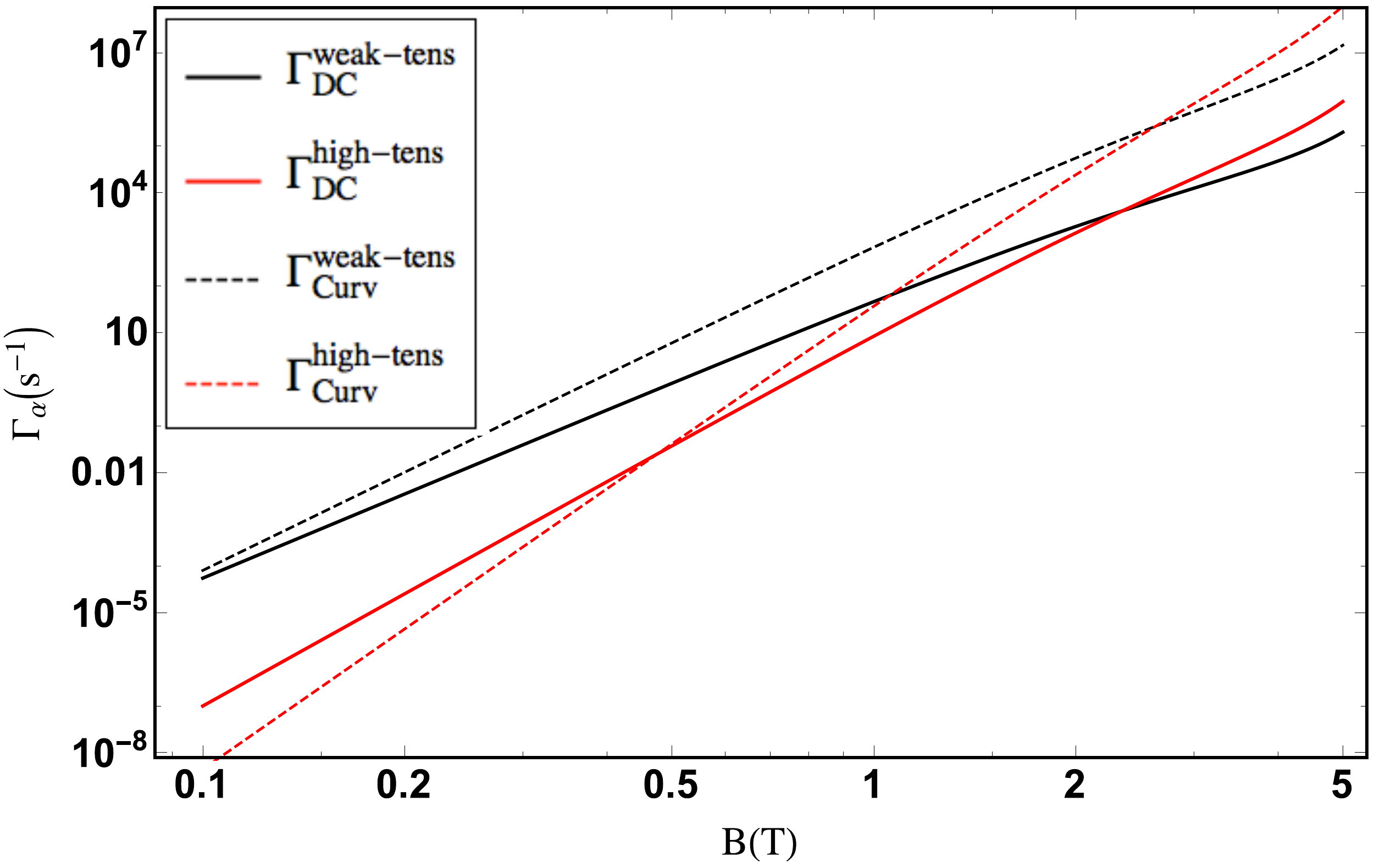}
			\caption{The relaxation rates for the direct spin-phonon coupling between Kramers pair states $(\Psi^{K}_{\uparrow},\Psi^{K'}_{\downarrow})$ as a function of perpendicular magnetic field. The calculation is preformed for a circular MoS$_2$ quantum dot with a radius of $R=50\textrm{nm}$ with hard wall boundary conditions and with a inter-valley coupling of $V_{KK'} = 0.1\textrm{meV}$. For both relaxation rates within the high tension regime we took the wave vector scale $q_{*}$ to be of the size such the largest energy scale in the problem is $\hbar \omega_{\mathbf{q}_*}$, and therefore we are always in regime where the entire phonon spectrum for the flexural modes are linear in $\mathbf{q}$.}	
\label{DirectSpinRelaxationRateFig}
\end{figure}

Therefore the Hamiltonian $H_{\textrm{DC}}$ describing spin-flip transition mediated by the deflection coupling mechanism is given by
\begin{equation} H_{\textrm{DC}} = -i\tau_z \lambda \zeta^{(F)}_{\mathbf{q}}( s_x q_x + s_y q_y ) (a^{(F)}_{\mathbf{q}} e^{i\mathbf{q}\cdot\mathbf{r}}+ a^{(F)\dagger}_{-\mathbf{q}} e^{-i\mathbf{q}\cdot\mathbf{r}}) \;.\end{equation}
Working with the dipole approximation we can now find the matrix element, this yields
\begin{equation} (H_{\textrm{DC}})_{nn}^{\downarrow\uparrow} = - i \frac{\lambda}{2} |\mathbf{q}| e^{-i\phi_\mathbf{q}}  \zeta^{(F)}_{\mathbf{q}} (\nu_{\downarrow}M^{KK}_{n,n} + \nu_{\uparrow}M^{K'K'}_{n,n}) \;, \end{equation}
where $M^{\tau,\tau'}_{n,n'} = \int dx x \chi_{n,s}^{\tau *}(x) \chi_{n',s}^{\tau'}(x)$. Finally we use Fermi's golden rule to compute the relaxation rate, as shown in Eq. (\ref{rateEq}). We first consider the case of weak tension where the phonon dispersion in quadratic, which yields  
\begin{equation} \Gamma_{\textrm{DC}}^{\textrm{weak-tens.}} = \frac{\lambda^2}{8\hbar \rho \beta^{2}_{\textrm{LT}}} \big| \nu_{\downarrow}M^{KK}_{n,n} + \nu_{\uparrow}M^{K'K'}_{n,n} \big|^2 \end{equation}
So we find that the relaxation rate only depends on the strength of the magnetic field by the orbital components of the matrix elements and on the valley mixing $\nu_s$. In Fig. \ref{DirectSpinRelaxationRateFig} we plot the relaxation rate $\Gamma_{\textrm{DC}}^{\textrm{weak-tens.}}$ and we see that while it does only vary in magnetic field by the matrix elements and valley mixing terms this can change my many orders of magnitude over a largest B-field range. While in the case of high tension, where the dispersion is linear we find  
\begin{equation}  \Gamma_{\textrm{DC}}^{\textrm{high-tens.}} = \frac{\lambda^2}{4\rho} \frac{\tilde{g}^2\mu_B^2B_z^2}{\hbar^3 \beta^{4}_{\textrm{HT}}} \big| \nu_{\downarrow}M^{KK}_{n,n} + \nu_{\uparrow}M^{K'K'}_{n,n} \big|^2 \;. \label{DCHightens}\end{equation}

The orbital electronic states in TMDCs are classified by their symmetry under mirror reflection ($z\rightarrow-z$), in the flat configuration the electronic states with different symmetries do not couple with each other. Local curvature of the TMDC lattice breaks this symmetry and allows for these previously forbidden spin non-conserving transitions between orbital electronic states. In a previous work we have developed a detailed model for the low-energy theory describing these effects.\cite{Pearce2016}

The Hamiltonian $H_{\textrm{curv}}$ for these processes is given by %arises from the the mixing of even and odd symmetry orbital electronic states and is given by 
\begin{equation} H_{\textrm{curv}} = -\tau_z (\xi_1 2 q_x q_y h_{\mathbf{q}} s_x + \xi_2 (q^2_x h_{\mathbf{q}} - q^2_y h_{\mathbf{q}}) s_y) \;,\end{equation}%+ s_z \delta \lambda_1 \textrm{Tr}[u_{ij}]
where for the case of MoS$_2$ the energy scales have been found to be $\xi_1=115\textrm{meV\AA}$ and $\xi_2=67\textrm{meV\AA}$. The matrix element within the dipole approximation is given by
\begin{align} (H_{\textrm{curv}})_{nn}^{\downarrow\uparrow} = &\; -\frac{1}{2} |\mathbf{q}|^2  \zeta^{(F)}_{\mathbf{q}}\big[\xi_1 \sin2\phi_{\mathbf{q}} + i\xi_2 \cos 2 \phi_{\mathbf{q}} \big] \nonumber \\ 
& \; \times (\nu_{\downarrow}M^{KK}_{n,n} + \nu_{\uparrow}M^{K'K'}_{n,n}) \;.\end{align}
We can once again calculate the relaxation rate for this direct spin-phonon coupling mechanism using Fermi's golden rule and we find, in the low tension regime
\begin{equation} \Gamma^{\textrm{weak-tens.}}_{\textrm{curv}} = \frac{\tilde{g}\mu_BB_z}{32\hbar^2\rho \beta^{3}_{\textrm{LT}}} (\xi_1^2 + \xi_2^2) \Big| \nu_{\downarrow}M^{KK}_{n,n} + \nu_{\uparrow}M^{K'K'}_{n,n} \Big|^2 \;. \end{equation} 
Whereas in the regime of high tension, we obtain
\begin{equation} \Gamma^{\textrm{high-tens.}}_{\textrm{curv}} = \frac{\tilde{g}^4\mu_B^4B_z^4}{16\hbar^5\rho \beta^{6}_{\textrm{HT}}} (\xi_1^2 + \xi_2^2)\Big| \nu_{\downarrow}M^{KK}_{n,n} + \nu_{\uparrow}M^{K'K'}_{n,n}\Big|^2 \;.\label{CurvHightens} \end{equation} 
We observe a higher power law dependance on the perpendicular magnetic field in the high tension regime within both spin-phonon coupling mechanisms. Due to the electrostatic gating structures necessary for the experimental realisation of a QD, it is expected that TMDs sheets would be most likely to be observed within the high tension regime. It is worth noting that these relaxation rates could be suppressed further if the sample is fabricated with large contact forces with the substrate, possibly due to encapsulation. This will induce a gapped flexural phonon dispersion which will close relaxation channels via flexural phonons with energies $\hbar\omega^{(F)}_{\mathbf{q}} < \Delta_{\textrm{Gap}}$, leading to reduced spin relaxation rates for B-fields below $B_z  >\Delta_{\textrm{Gap}}/\tilde{g}\mu_B$.

In Fig. \ref{DirectSpinRelaxationRateFig} we present the relaxation rates obtained in this section. We see that due to the large intrinsic spin orbit coupling in the TMDs the role of the direct spin-phonon coupling mechanism is very important for understand all spin relaxation mechanisms. In the high tension regime we see that at low B-fields the relaxation rates are suppressed quite significantly, while at high B-fields will become the dominant relaxation mechanism for spin relaxation due the increased power of $B_z$ appearing in the rates seen in Eq. (\ref{DCHightens}) and Eq. (\ref{CurvHightens}).    

%Level crossings at finite a B-field offer one platform for creating electronic states which can be used as qubits. 
%The states will be orbital states of different orbital quantum numbers, opposite spin and which belong to the same valley. In Sec. III.A we consider the indirect admixture mechanism where the electron spin relaxes due to either a piezoelectric phonon or a deformation potential phonon and then in Sec. III.B we explore the role of direct spin-phonon coupling.
\subsection{Pure Spin or Valley Relaxation}

The energy spectrum of TMD QD under a perpendicular magnetic field will exhibit several level crossings as the magnetic field is increased. These level crossings can be used as pure valley or spin qubit. In this section we will consider the relaxation rates between these pure spin and valley states meditated by the admixture mechanism.

For the lowest orbital levels of the QD high B-field of magnitude $B>5\textrm{T}$ are required for level crossings between states with the same angular quantum number $l$, as shown in Fig. \ref{DotLevelsFig}.c. By considering orbital states with large values of $l$ one can find level crossings at lower magnitudes of magnetic field, but this introduces additional orbital relaxation effects which can occur on very fast time scales. As a consequence we will focus our attention on level crossings in the lowest orbital level. For concreteness we consider eigenstates within the same valley but of opposite spins which have a level crossing at a finite B-field $B_0$, and study the relaxation of the spin within the same valley and orbital level. The energy spectrum around the level crossing can be linearised around the B-field $B_0$, this is a good approximation given that $g_e\mu_B[B_z-B_0]/(E_n-E_k)\ll1$.

As this spin relaxation can occur between states of the same valley there is an effective time reversal symmetry breaking, as the time reversed partner state to any state within one valley will belong to another inequivalent valley.\cite{Struck2010} As a consequence their will be no Van Vleck cancelation in this regime, leading to a lower dependance of the spin relaxation matrix element on the perpendicular magnetic field.

Following the same procedure for calculating the relaxation rate for the admixture mechanism outlined in Sec. III.A we find the relaxation rates for the case of pure spin relaxation. This process will yield, 
\begin{align} & \Gamma^{KK}_{\textrm{L},\textrm{DP}} = \; \frac{16\pi^4\lambda^2_{R}g_1^2}{\rho v^{(L)6}\hbar^5} [g_e\mu_B(B_z-B_0)]^4 \nonumber \\
& \times \Bigg| \sum_{k\neq n} \frac{\delta_{n,k-1}N^{K,K}_{n,k}(k_K)^{K,K}_{k,n}}{E^K_n - E^K_k - 2\lambda} + \frac{\delta_{n,k+1}(k_K)^{K,K}_{n,k}N^{K,K}_{k,n}}{E^K_n - E^K_k + 2\lambda} \Bigg|^2 \;. \label{PureSpin}\end{align}
The relaxation rates due to coupling with piezoelectric phonons can be found from the same expressions as given in Eq. (\ref{RelaxPiezo2}) and Eq. (\ref{RelaxPiezo}), while spin relaxation rates for the K$'$ are acquired via the interchange to valley indices in Eq. (\ref{PureSpin}). The starkest difference to the result found in Eq. (\ref{RelaxDefPot}) is that the matrix element is no longer surpassed by a factor of ratio Zeeman energy and phonon energy due to Van Vleck cancelation. This result gives an analogous result as has been predicted for graphene, but with the difference that the energy scale of the spin-orbit splitting is larger than the orbital level splitting due to confinement.  

\section{Conclusions}

In this work we have investigated the electron spin relaxation mechanisms and their respective $T_1$ times in TMDs QDs. In the low field regime where we consider the relaxation within a Kramers doublet added by disorder induced valley mixing, as this seems the most achievable experimental situation. We find the magnetic field dependance as $\propto B^6$ due to the role of the valleys giving rise to Van Vleck cancelation and the nature of the two-dimensional phonon properties, also found is a non-monotonic behaviour arising to a destructive valley interference. We also find a relaxation due to direct spin-phonon coupling which does not depend of B-field to lowest order, but is also strongly dependant on tension within the sample. 

We see that due the combined spin-valley structure and large intrinsic spin orbit coupling the TMDs provides new platform for spintronic applications. In the future it would be important to explore mechanisms of control of qubits in TMDs, particularly it would be of interest to study how the spin-phonon coupling can be manipulation for greater control of the electronic spin. 

\acknowledgements

We thank A. Korm\'{a}nyos and M. Brooks for fruitful discussions. We acknowledge funding from DFG under the program SFB 767 and the EU through Marie Curie ITN Spin-Nano.

\appendix

\section{The Momentum Operator}

In this appendix we will present the explicit relations for the momentum operator acting upon the eigenfunctions of the QD presented in Sec. II. These eigenfunctions are given by $\Psi^{\tau}_s(x,\varphi) = e^{il\varphi}\chi_{l,s}^{\tau}(x)$ where we take $\chi_{l,s}^{\tau}(x) = x^{\frac{|l|}{2}} e^{-\frac{x}{2}} M(a_{l,\tau}^{<},|l|+1,x)$. Taking the definition of the momentum operator presented in Sec II. and using confluent hypergeometric function identities\cite{Kormanyos2014,AbraStegun} we obtain, for the case that $l\geq0$,
\begin{align} k_{K}e^{il\varphi}\chi_{l,s}^{K}(x) = &\; \frac{i\sqrt{2}}{l_B}e^{i(l+1)\varphi}\big[1-\frac{a_{l,K}^{<}}{l+1}\big]\chi_{l+1,s}^{K}(x) \;, \\ 
k_{K'}e^{il\varphi}\chi_{l,s}^{K'}(x) = &\; \frac{i\sqrt{2}}{l_B}e^{i(l-1)\varphi}l\chi_{l-1,s}^{K'}(x) \;,
%k^{\dagger}_{K}e^{il\varphi}\chi_{l,s}^{K}(x) = &\; -\frac{i\sqrt{2}}{l_B}e^{i(l-1)\varphi}l\chi_{l-1,s}^{K'}(x) \\
%k^{\dagger}_{K'}e^{il\varphi}\chi_{l,s}^{K'}(x) = &\;-  \frac{i\sqrt{2}}{l_B}e^{i(l+1)\varphi}\big[1-\frac{a_{l,K'}^{<}}{l+1}\big]\chi_{l+1,s}^{K'}(x)
\end{align}
whereas, for the case in which $l<0$,
\begin{align} k_{K}e^{il\varphi}\chi_{l,s}^{K}(x) = &\; - \frac{i\sqrt{2}}{l_B}e^{-i(l-1)\varphi}l\chi_{l-1,s}^{K}(x) \;, \\ 
k_{K'}e^{il\varphi}\chi_{l,s}^{K'}(x) = &\; \frac{i\sqrt{2}}{l_B}e^{-i(l+1)\varphi}\frac{a_{l,K'}^{<}}{l+1}\chi_{l+1,s}^{K'}(x) \;. \end{align}
We see that these act as raising and lowering operators on the angular momentum. Expressions for the action of the operator $k^{\dagger}_{\tau}$ upon the eigenfunctions, can easily be obtained using the relation $k^{\dagger}_{K} = - k_{K'}$ as these operators are related by time reversal symmetry.


\begin{thebibliography}{99}

\bibitem{Wang2012} Q.H.\ Wang, K.\ Kalantar-Zadeh, A.\ Kis, J.N. Coleman and M.S.\ Strano, Nat. Nanotechnol. {\bf7}, 699 (2012).

\bibitem{Mak2010} K.F.\ Mak, C.\ Lee, J.\ Horne, J.\ Shan and T.F. Heinz, Phys. Rev. Lett. {\bf105}, 136805 (2010).

\bibitem{Zhu2011} Z.Y.\ Zhu, Y.C.\ Cheng and U.\ Schwingenschl\"{o}gl, Phys. Rev. B. {\bf84}, 153402 (2011).

\bibitem{Kormanyos2015} A.\ Korm\'{a}nyos, G.\ Burkard, M.\ Gmitra, J.\ Fabian, V.\ Z\'{o}lyomi, N.D.\ Drummond and V.\ Fal'ko, 2D Mater. {\bf2}, 022001 (2015).

\bibitem{Xiao2012} D.\ Xiao, G-B.\ Liu, W.\ Feng, X.\ Xu, and W.\ Yao, Phys. Rev. Lett. {\bf108}, 196802 (2012).

\bibitem{Cao2012} T.\ Cao, G.\ Wang, W.\ Han, H.\ Ye, C.\ Zhu, J.\ Shi, Q.\ Niu, P.\ Tan, E.\ Wang, B.\ Liu and Ji.\ Feng, Nat. Commun. {\bf3}, 887 (2012).

\bibitem{Mak2010a} K.F.\ Mak, K.\ He, J.\ Shan and T.F.\ Heinz, Nat. Nanotechnol. {\bf7}, 494 (2012). 

\bibitem{Jones2013} A.M.\ Jones, H.\ Yu, N.J.\ Ghimire, S.\ Wu, G.\ Aivazian, J.S.\ Ross, B.\ Zhao, J.\ Yan, D.G.\ Mandrus, D.\ Xiao, W.\ Yao and X.\ Xu, Nat. Nanotechnol. {\bf8} 634 (2013). 

\bibitem{Loss1998} D.\ Loss and D.P.\ DiVincenzo, Phys. Rev. A. {\bf57} 120 (1998).

\bibitem{Hanson2007} R.\ Hanson, L.P.\ Kouwenhoven, J.R.\ Petta, S.\ Tarucha and L.M.K.\ Vandersypen, Rev. Mod. Phys. {\b79}, 1217 (2007).

\bibitem{Rohling2012} N.\ Rohling and G.\ Burkard, New Journal of Physics {\bf14}, 083008 (2012). 

\bibitem{Huang2014} C,\ Huang, S.\ Wu, A.M.\ Sanchez, J.J.P.\ Peters, R.\ Beanland, J.S.\ Ross, P.\ Rivera, W.\ Yao, D.H.\ Cobden and X.\ Xu, Nat. Materials {\bf13} 1096 (2014).

\bibitem{Song2015a} X-X.\ Song, Z-Z.\ Zhang, J.\ You, D.\ Liu, H-O.\ Li, G.\ Cao, M.\ Xiao and G-P.\ Guo, Scientific Reports {\bf5}, 16113 (2015). 

\bibitem{Song2015b} X-X.\ Song, Di.\ Liu, V.\ Mosallanejad, J.\ You, T-Y.\ Han, D-T.\ Chen, H-O.\ Li, G.\ Cao, M.\ Xiao, G-C.\ Guoa and G-P.\ Guo, Nanoscale {\bf7}, 16867 (2015). 

\bibitem{Wang2016} K.\ Wang, T.\ Taniguchi, K.\ Watanabe and P.\ Kim, arXiv:1610.02929 

\bibitem{Lee2016} K.\ Lee, G.\ Kulkarni and Z.\ Zhong, Nanoscale {\bf8}, 7755 (2016).

\bibitem{Kormanyos2014} A.\ Korm\'{a}nyos, V.\ Z\'{o}lyomi, N.D.\ Drummond and G.\ Burkard, Phys. Rev. X. {\bf4} 011034 (2014).

\bibitem{Liu2014} G-B.\ Liu, H.\ Pang, Y.\ Yao and W.\ Yao, New Journal of Physics {\bf16}, 105011 (2014).

\bibitem{Wu2016} Y.\ Wu, Q.\ Tong, G-B.\ Liu, H.\ Yu and W.\ Yao, Phys. Rev. B. {\bf93}, 045313 (2016). 

\bibitem{Tonndorf2015} P.\ Tonndorf, R. Schmidt, R.\ Schneider, J.\ Kern, M.\ Buscema, G.A.\ Steele, A.\ Castellanos-Gomez, H.S.J.\ van der Zant, S.Michaelis de \ Vasconcellos and R.\ Bratschitsch, Optica {\bf}, 347 (2015).

\bibitem{Srivastava2015a} A.\ Srivastava, M.\ Sidler, A.V.\ Allain, D.S. Lembke, A.\ Kris and A.\ Imamo\~{g}lu, Nat. Nanotechnol. {\bf10}, 491 (2015). 

\bibitem{He2015} Y-M.\ He, G.\ Clark, J.R. Schaibley, Y.\ He M-C.\ Chen, Y-J.\ Wei, X.\ Ding, Q.\ Zhang, W.\ Yao, X.\ Xu, C-Y.\ Lu and J-W.\ Pan, Nat. Nanotechnol. {\bf10}, 497 (2015). 

\bibitem{Koperski2015} M.\ Koperski, K.\ Nogajewski, A.\ Arora, V.\ Cherkez, P.\ Mallet, J-Y.\ Veuillen, J.\ Marcus, P.\ Kossacki and M.\ Potemski, Nat. Nanotechnol. {\bf10}, 503 (2015).   

\bibitem{Chakraborty2015} C.\ Chakraborty, L.\ Kinnischtzke, K.M. Goodfellow, R.\ Beams and A.N.\ Vamivakas, Nat. Nanotechnol. {\bf10}, 507 (2015). 

\bibitem{Kaasbjerg2012} K.\ Kaasjberg, K.S.\ Thygesen and K.W. Jacobsen, Phys. Rev. B. {\bf85}, 115317 (2012).

\bibitem{Kaasbjerg2013} K.\ Kaasjberg, K.S.\ Thygesen and A-P.\ Jauho, Phys. Rev. B. {\bf87}, 235312 (2013).

\bibitem{khaetskii2000} A.V.\ Khaetskii and Y.V.\ Nazarov, {\bf61}, 12639 (2000).

\bibitem{khaetskii2001} A.V.\ Khaetskii and Y.V.\ Nazarov, {\bf64}, 125316 (2001).

\bibitem{Pearce2016} A.J.\ Pearce, E.\ Mariani and G. Burkard, Phys. Rev. B. {\bf94}, 155416 (2016).

\bibitem{Schmidt2016} H.\ Schmidt, I.\ Yudhistira, L.\ Chu, A.H.\ Castro Neto, B.\ \"{O}zyilmaz, S.\ Adam and G.\ Eda, Phys. Rev. Lett. {\bf116}, 046803 (2016).

\bibitem{Srivastava2015} A.\ Srivastava, M.\ Sidler, A.V.\ Allain, D.S.\ Lembke, A.\ Kis and A.\ Imamo\v{g}lu, Nat. Phys. {\bf11} 141-147 (2015). 

\bibitem{Aivazian2015} G.\ Aivazian, Z.\ Gong, A.M.\ Jones, R-L.\ Chu, J.\ Yan, D.G.\ Mandrus, C.\ Zhang, D.\ Cobden, W.\ Yao and X.\ Xu,  Nat. Phys. {\bf11} 148-152 (2015). 

\bibitem{MacNeill2015} D.\ MacNeill, C.\ Heikes, K.F.\ Mak, Z.\ Anderson, A.\ Korm\'{a}nyos, V.\ Z\'{o}lyomi, J.\ Park and D.C.\ Ralph, Phys. Rev. Lett. {\bf114}, 037401 (2015).

\bibitem{Dias2016} A.C.\ Dias, J.\ Fu, L.\ Villegas-Lelovsky and F.\ Qu, J. Phys.: Condens. Matter {\bf28}, 375803 (2016). 

\bibitem{AbraStegun} {\it Handbook of Mathematical Functions}, edited by M.\ Abramowitz and I.A.\ Stegun, Dover, New York, (1965).

\bibitem{Recher2009} P.\ Recher, J.\ Nilsson, G.\ Burkard and B.\ Trauzettel, Phys. Rev. B {\bf79}, 085407 (2009).

\bibitem{Palyi2011} A.\ P\'{a}lyi and G.\ Burkard, Phys. Rev. Lett. {\bf106}, 086801 (2011).

\bibitem{Tsitsishvili2004} E.\ Tsitsishvili, G.S.\ Lozano and A.O.\ Gogolin, Phys. Rev. B. {\bf70}, 115316 (2004).

\bibitem{Bulaev2008} D.V.\ Bulaev, B.\ Trauzettel and D.\ Loss, Phys. Rev. B. {\bf77}, 235301 (2008).

\bibitem{Struck2010} P.R.\ Struck and G.\ Burkard, Phys. Rev. B. {\bf82}, 125401 (2010).

\bibitem{Rudner2010} M.S.\ Rudner and E.I.\ Rashba, Phys. Rev. B. {\bf81}, 125426 (2010).  

\bibitem{Baugher2013} B.W.H.\ Baugher, H.O.H.\ Churchill, Y.\ Yang and P.\ Jarillo-Herrero Nano Lett. {\bf13}, 4212-4216 (2013)

\bibitem{Yang2015} L.\ Yang, N.A.\ Sinitsyn, W.\ Chen, J.\ Yuan, J.\ Zhang, J.\ Lou and S.A. Crooker, Nat. Phys. {\bf11}, 830-834 (2015).

\bibitem{Zhou2013} W.\ Zhou, X.\ Zou, S.\ Najmaei, Z.\ Liu, Y.\ Shi, J.\ Kong, J.\ Lou, P.M.\ Ajayan, B.I.\ Yakobson and J-C.\ Idrobo, Nano Lett. {\bf13}, 2615-2622 (2013).

\bibitem{Hong2015} J.\ Hong, Z.\ Hu, M.\ Probert, K.\ Li, D.\ Lv, X.\ Yang, L.\ Gu, N.\ Mao, Q.\ Feng, L.\ Xie, J.\ Zhang, D.\ Wu, Z.\ Zhang, C.\ Jin, W.\ Ji, X.\ Zhang, J.\ Yuan and Z.\ Zhang, Nat. Commun. {\bf6}, 6293 (2015).

\bibitem{Lin2015} Y-C.\ Lin, T.\ Bj\"{o}rkman, H-P.\ Komsa, P-Y.\ Teng, C-H.\ Yeh, F-S.\ Huang, K-H.\ Lin, J.\ Jadczak, Y-S.\ Huang, P-W.\ Chiu, A.V.\ Krasheninnikov and K.\ Suenaga, Nat. Commun. {\bf6}, 6736 (2015).

\bibitem{Ochoa2014} H.\ Ochoa, F.\ Finocchiaro, F.\ Guinea and V.I.\ Fal'ko, Phys. Rev. B. {\bf90}, 235429 (2014).

\bibitem{Habe2016} T.\ Habe and M.\ Koshino, Phys. Rev. B. {\bf93}, 075415 (2016).

\bibitem{Ando1998} T.\ Ando and T.\ Nakanishi, J. Phys. Soc. Jpn. {\bf67} 1704 (1998).

\bibitem{Palyi2010} A.\ P\'{a}lyi and G.\ Burkard, Phys. Rev. B. {\bf82}, 155424 (2010).

\bibitem{Vleck1940} J.H.\ Van Vleck, Phys. Rev. {\bf57}, 426 (1940).

\bibitem{Abrahams1957} E.\ Abrahams, Phys. Rev. {\bf107}, 491 (1957).

\bibitem{Jiang2013} J-W.\ Jiang, Z.\ Qi, H.S.\ Park and T.\ Rabczuk, Nanotechnology {\bf24}, 435705 (2013). 

\end{thebibliography}
\end{document}